# Computing Correct Truncated Excited State Wavefunctions

N.C. Bacalis[1, a)] Z. Xiong[2, b)], J. Zang[2] and D. Karaoulanis[3]

[1]*Theoretical and Physical Chemistry Institute, National Hellenic Research Foundation, Athens, GR-11635, Greece.*
[2]*Space Science and Technology Research Institute, Southeast University, Nanjing, 21006, Peoples Republic of China*
[3]*Korai 21, Halandri, GR-15233, Greece.*

[a)]Corresponding author: nbacalis@eie.gr
[b)]zhuangx@seu.edu.cn

**Abstract.** We demonstrate that, if a wave function's truncated expansion is small, then the standard excited states computational method, of optimizing one "root" of a secular equation, may lead to an incorrect wave function - despite the correct energy according to the theorem of Hylleraas, Undheim and McDonald - whereas our proposed method [J. Comput. Meth. Sci. Eng. **8**, 277 (2008)] (independent of orthogonality to lower lying approximants) leads to correct reliable small truncated wave functions. The demonstration is done in He excited states, using truncated series expansions in Hylleraas coordinates, as well as standard configuration-interaction truncated expansions.

## INTRODUCTION

Using large wave function expansions in truncated - but as complete as possible - spaces, is generally safe, but is rather impracticable, especially when dealing with large systems. Generally, having a small and handy, but reliable, expansion is much preferable if it is "useful" i.e. if it curries the main properties of the system, while any corrections aim in improving the energy by describing the "splitting" of the wave function near the nuclei (due to the strong electron repulsion there, and to the Pauli principle). Obtaining such a "useful" wave function for the ground state is relatively easy by minimizing the energy, but for excited states, minimization of the energy can only be achieved if the wave function is orthogonal to all lower states. But this requires accurate large expansions. Here we demonstrate that if the truncated lower lying wave functions are small, then (habitually) orthogonalizing to them may lead to disastrous results, although the energy may tend to the correct value according to the theorem of Hylleraas, Undheim and McDonald (HUM) [1]. On the contrary, by minimizing our proposed functional Ω [2], a correct wave function is obtained, although "small", currying the same main properties as the "large" function (obtained comparable, and safely, by either HUM or Ω). Note that Ω does **not** need orthogonalization to lower lying wave functions; orthogonality should be an outcome. The demonstration is done in He excited states, using truncated series expansions in Hylleraas coordinates, as well as standard configuration-interaction (CI) truncated expansions.

## The excited state energy $E_n$ is a saddle point

Expand the approximant $|\phi_n\rangle$ around the exact state $|n\rangle$ (assumed real and normalized) where $E_0 < E_1 < E_2 \cdots$, write the energy as $\langle \phi_n | H | \phi_n \rangle = E = -L + E_n + U$ where, in terms of the coefficients, the lower, $L$, and the higher, $U$, terms (saddle) are:   ***down-paraboloids***   and   ***up-paraboloids*** :

$$E = \boxed{-L = -\sum_{i<n}(E_n - E_i)\langle i|\phi_n\rangle^2} + E_n + \boxed{U = \sum_{i>n}(E_i - E_n)\langle i|\phi_n\rangle^2} \quad (1)$$

If $n=0$, or if $\phi_n$ is orthogonal to all lower $|\psi_i\rangle = |i\rangle$ (which can be approximated satisfactorily by $\phi_i$ only if $\phi_i$ are "large" expansions), then $L$ is absent, and minimizing $E=E_n+U$ is sufficient. But if $\phi_i$ are "small" expansions (not accidentally orthogonal to $\phi_n$), then $L$ is present. Minimizing $E=-L+E_n+U$ "orthogonally to all lower $\phi_i$", must lead to $E$ below $E_n$, because (consider e.g. the 1st excited state) $\phi_1^+$, the **closest** function to $\psi_1$ **while** orthogonal to $\phi_0$, lies **below** $E_1$ (and the minimum is even lower). The HUM theorem demands $E[\phi_n] > E_n$ *while* $\phi_n$ *is* orthogonal to all lower "roots" $\phi_i$ of the secular equation. Therefore, $\phi_1^+$ is not accessible by HUM (and even more inaccessible is $\psi_1$). Even worse, note that in optimizing any HUM root (say $\phi_1$), all other roots ($\phi_0, \phi_2, ...$) get deteriorated since we may have $\langle 1|\phi_1\rangle^2 \to 1$, *at will, but*: $\langle 0|\phi_0\rangle^2 + \langle 0|\phi_1\rangle^2 + \langle 0|\phi_2\rangle^2 + \cdots + \langle 0|\phi_N\rangle^2 \leq 1 \Rightarrow \langle 0|\phi_0\rangle^2 < 1 - \langle 0|\phi_1\rangle^2$.

Therefore, the optimized HUM 2nd root $\phi_1$ (although $E[\phi_1] > E_1 > E[\phi_1^+]$) is orthogonal to a **deteriorated** 1st root $\phi_0$, i.e. $\phi_1$ just stops at $E_1$ and cannot approach a **worse** $\phi_1^+$ (the **closest** to $\psi_1$ while orthogonal to the **deteriorated** $\phi_0$), thus, **$\phi_1$ is much more veered away from the exact** $\psi_1$. This is clearly demonstrated below for He. (If the optimized HUM roots are misleading for He, i.e. the smallest atom, **there is no guarantee for larger systems!**)

## TOOLS AND APPROXIMATIONS

We need very accurate (truncated) functions $\Psi_n$ to resemble $\approx$ eigenfunctions $\psi_n$ and truncated approximants $\Phi_n$ to check the closeness to $\Psi_n$. As truncated functions we use

1. For He $^1S$ ($1s^2$ and $1s2s$): **Series expansion in Hylleraas variables** $s = r_1 + r_2$, $t = r_1 - r_2$, $u = |\vec{r}_1 - \vec{r}_2|$. $\Phi(r_1,r_2)$ consist of one **Slater determinant** of (non-linear) variational Laguerre–type orbitals, $1s$, $2s$ **multiplied** by a truncated power series of $s,t,u$, as (linear) eigenvectors of 1st or 2nd root of a secular equation. For the "exact" $\Psi_n$ we go up to **27 terms**, $E_0 \approx -2.90371$ a.u., $E_1 \approx -2.14584$ a.u., compared to Pekeris' 95 terms: $E_0 = -2.90372$, $E_1 = -2.14597$ a.u. [3]. For the "truncated" trial functions $\Phi_n$ we go up to **8 terms.**

2. For He $^1S$ ($1s^2$, $1s2s$ and $1s3s$) and for He $^3S$ ($1s2s$ and $1s3s$) we use **Configuration Interaction (CI) in spherical coordinates $(r,\theta,\varphi)$**. $\Phi(r_1,r_2)$ is a linear combination of configurations out of Slater determinants (SD) of atomic (non-linear) variationally optimized Laguerre-type spin-orbitals (orthogonalized) and the linear CI coefficients are the eigenvectors of the roots of the secular equation. As "exact" $\Psi_n$ we use a "large" expansion in **1s, 2s, 3s, 4s, 5s, 2p, 3p, 4p, 5p, 3d, 4d, 5d, 4f, 5f**. $^1S$: $E_0 \approx -2.90324$ a.u., $E_1 \approx -2.14594$ a.u., $E_2 \approx -2.06125$ a.u. (exact: -2.06127 a.u. [3]), $^3S$: $E_0 \approx -2.17521$ a.u., $E_1 \approx -2.06869$ a.u. (exact: -2.17536, -2.06881 a.u. [3]). As "truncated" trial functions $\Phi_n$ we use a "small" expansion in **1s, 2s, 3s**.

## TWO METHODS

We shall use **two methods**:

1. Minimizing (optimizing) directly the **$n^{th}$ HUM root**, which, [cf. above], must be veered away from the exact eigenfunction $\psi_n$, because, according to the HUM theorem it tends to the exact energy from above, unable to get closer to the exact $\psi_n$, that would require taking lower energies, due to orthogonality to deteriorated lower roots.

2. Minimizing the functional $\Omega_v$ that has **minimum at** the **exact $\psi_n$** [2]

$$\Omega_n[\phi_0,\phi_1,...;\phi_n] \equiv E[\phi_n] + 2\sum_{i<n} \frac{\langle \phi_i | H - E[\phi_n] | \phi_n \rangle^2}{E[\phi_n] - E[\phi_i]} \left[1 - \sum_{i<n} \langle \phi_i | \phi_n \rangle^2\right]^{-1} \quad (2)$$

obtained by inverting the sign of $L$ (the down parabolas): $E = +L + E_n + U$. The lower $\phi_i$ may be **inaccurate and very "small"**, provided that the Hessian and all its principal minors along the main diagonal be positive, easy to fulfill because their main term is large and the overlaps in 1+2[...] (c.f. ref. [2]) are small.

## RESULTS

The main orbitals of the Hylleraas wave functions for **He $^1S$ $1s2s$** are shown in Fig. 6 of ref [4]. Clearly, the HUM-wave function is **not $1s$"$2s$"**! (It has a "node" at 10 a.u.! So, it is essentially $1s1s'$). The whole wave function needs **8** Hylleraas series **terms** to be fixed (or even **27** terms!). On the contrary, the $\Omega$- wave function (both the "large", of 27 terms, and the "small", of 8 terms, are correct and **practically identical**. This was expected because $Ω_1$ has minimum **at** the exact (saddle point) 1s2s, whereas the HUM 2$^{nd}$ root is orthogonal to a necessarily deteriorated 1$^{st}$ root, therefore is veered away from the exact 1s2s.

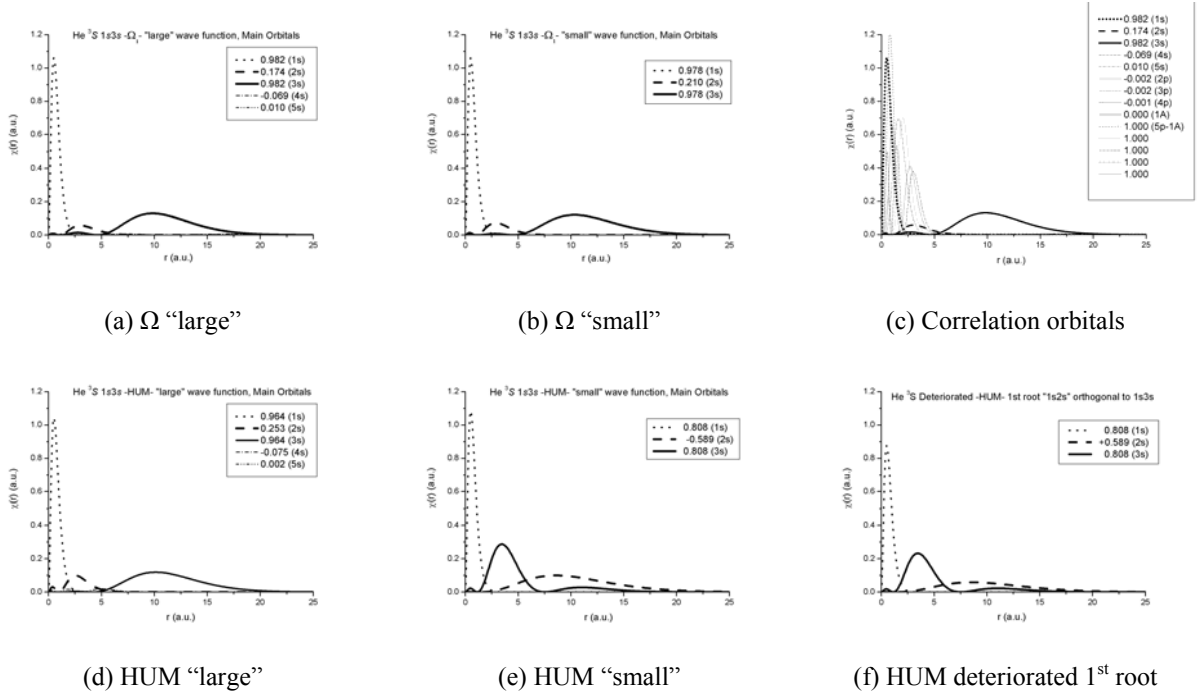

| (a) Ω "large" | (b) Ω "small" | (c) Correlation orbitals |
| (d) HUM "large" | (e) HUM "small" | (f) HUM deteriorated 1st root |

**FIGURE 1.** CI wave functions (main orbitals): for **He $^3S$ 1s3s**. Solid: 3s, Dashed: 2s, Dotted: 1s.

The CI wave functions (main orbitals), HUM and Ω, are compared for **He $^3S$ 1s3s** in Fig. 1. The "large" functions Ω (a) and HUM (d) are **almost identical**, and they have the same main orbitals as the Ω "small" function (b), namely 1s and 3s, where the 2s adds some correlation correction near the nucleus (as well as all higher orbitals of the "large" expansion (c)). But HUM "small" expansion (e), orthogonal to a deteriorated 1$^{st}$ root "1s2s" (f), **has main orbitals 1s2s** (with opposite sign), trying to correct the total wave function approaching the correct energy, by using the 3s orbital for correlation correction. This **is misleading** because it proposes to the audience, as "HOMO" orbital, the 2s instead of the 3s. Similar results are obtained also for He $^3S$ 1s3s: Ω "small" is, correctly, mainly 1s3s, whereas HUM "small" is, **misleadingly**, mainly 1s2s, instead of 1s3s. For He $^1S$ 1s2s the "small" HUM and Ω functions are the essentially the same.

## Reliability Criteria

Since the **exact $E_n$** is a **saddle point** (cf. Fig. 5 of ref [4]), if we stop the Ω minimization with some convergence criterion and the energy happens to be either **slightly <u>higher</u>**, or **slightly <u>lower</u>**, is the "small" wave function reliable? There are two criteria: (i) By changing slightly several parameters we must check that the final point (minimum of Ω) is saddle in energy $E$. (ii) This will never happen *exactly*, so we must check that the difference (Ω-$E$) is small. If a "large" function is available (to serve as "exact"), before we release the "small" function to the audience, there is a third criterion: In $E=-L+E_n+U$ [cf. Eq. 1], the unknown $U$ is $U≥0$. Then **$E ≥ -L+E_n$**. In fact this is **the correct lower bound** of $E$, and **not just $E_n$**. $E_n$ would be a lower bound of $E[\phi_n] > E_n$ if either (a) $\phi_n$ were exactly orthogonal to all lower eigenfunctions (which **never** happens and is approximately fulfilled if the functions are "large") or (b) if $\phi_n$ were the $(n+1)^{th}$ HUM root (which is **always veered away** from the exact eigenfunction and approaches it also if it is "large"). But Shull and Löwdin [5] have shown that any excited state can be computed independently of the lower lying approximants, and this exactly, is done by Ω. Therefore, the **correct** lower bound is $E=-L+E_n$, so that the 3$^{rd}$ reliability criterion is that (iii) $L$ should be small and $0 < E_n - E ≤ L$. Using as $\psi_0$ our "large" wave functions $\phi_0$, we have the following Table 1.

**TABLE 1.** Estimating the 3rd reliability criterion.

| Wave function | Ω | E | Ω-E | $E_n$ | $E - E_n$ | L | $L > E_n - E$ ? |
|---|---|---|---|---|---|---|---|
| $^1S$ 1s2s Large | -2.145934 | -2.14594 | $2\ 10^{-9}$ | -2.14597 | $3\ 10^{-5}$ | $2\ 10^{-9}$ | |
| $^1S$ 1s3s Large | -2.061252 | -2.061252 | $3\ 10^{-13}$ | -2.06127 | $2\ 10^{-5}$ | $3\ 10^{-8}$ | |
| $^1S$ 1s3s Small | -2.049335 | -2.06278 | $1.3\ 10^{-3}$ | -2.06127 | -0.002 | 0.006 | YES |
| $^3S$ 1s3s Large | -2.06835 | -2.06835 | $1\ 10^{-14}$ | -2.06881 | $5\ 10^{-4}$ | $5\ 10^{-6}$ | |
| $^3S$ 1s3s Small | -2.06795 | -2.06934 | $1.4\ 10^{-3}$ | -2.06881 | $-5.3\ 10^{-4}$ | $6\ 10^{-4}$ | YES |

We have two cases lying **below** the exact. But $\Omega - E$ is small, $L$ is small (and $L > E_{exact} - E$), so, they are reliable.

## INCORRECT FUNCTIONS WITH CORRECT ENERGY

The correct energy is not a safe criterion of correctness. Infinitely many functions $\Phi$ **orthogonal to $\psi_1$** can have energy $E[\Phi] = E[\psi_1]$: Take any normalized function $\Psi$, and $\Psi_\perp$ orthogonal to $\psi_0$ and $\psi_1$: Then the function

$$\Phi = \sqrt{\frac{E[\Psi_\perp] - E_1}{E[\Psi_\perp] - E_0}}\psi_0 + 0\psi_1 - \sqrt{\frac{E_1 - E_0}{E[\Psi_\perp] - E_0}}\Psi_\perp, \quad \left(\Psi_\perp = \frac{\Psi - \psi_0\langle\psi_0|\Psi\rangle - \psi_1\langle\psi_1|\Psi\rangle}{\sqrt{1 - \langle\psi_1|\Psi\rangle^2 - \langle\psi_\perp|\Psi\rangle^2}}\right) \quad (3)$$

has energy $E[\Phi] = E[\psi_1] = E_1$ and is orthogonal to $\psi_1$. Some examples are shown in Table 2. Starting from a state $\Psi$ of energy $E[\Psi]$ we found a function $\Phi = A\,\psi_0 + B\,\psi_1 + C\,\Psi$, such that: $\langle\psi_1|\Phi\rangle \sim 0$, and $\langle\Phi|H|\Phi\rangle = E[\psi_1] \sim E_1$.

**TABLE 2.** Linear combinations, $\Phi$, of higher eigenfunctions with $\psi_0$, which are orthogonal to $\psi_1$, but have energy $E[\Phi] = E[\psi_1] = E_1$.

| $E[\Psi]$ | A | B | C | $\langle\psi_1|\Phi\rangle$ | $\langle\Phi|H|\Phi\rangle$ | $\psi_1$ | $E_1$ |
|---|---|---|---|---|---|---|---|
| -2.01990 | 0.37769 | 0.00004 | -0.92591 | -0.00002 | -2.14594 | $^1S$ | -2.14594 |
| -2.03650 | 0.47559 | -0.00012 | -0.87635 | 0.00006 | -2.06852 | $^3S$ | -2.06868 |
| -2.01990 | 0.58239 | 0.00020 | -0.81298 | -0.00040 | -2.05512 | $^1P$ | -2.05512 |

Of course, these contain remote electrons and cannot be used as approximants of $\psi_1$, since they are orthogonal to $\psi_1$. But if entanglement is experimentally achieved, it might be possible to accomplish a reaction at $E_1$ via more remote electrons.

## ACKNOWLEDGEMENTS

This work was sponsored by: Key Project of National Social Science (Grant No.15AJL004), China, Polynano-Kripis 447963 / GSRT, Greece.